# Controllable Lateral Optical Forces on Janus Particles in Fluid Media

Ziheng Xiu[1], Yue Chai[1], Pengyu Wen[1], Chun Meng[2], Min-Cheng Zhong[2], Yu-Xuan Ren[3], Daohong Song[1], Hrvoje Buljan[1,4], Liqin Tang[1,5]* and Zhigang Chen[1,5]*

[1]*The MOE Key Laboratory of Weak-Light Nonlinear Photonics, TEDA Applied Physics Institute and School of Physics, Nankai University, Tianjin 300457, China*

[2]*Anhui Province Key Laboratory of Measuring Theory and Precision Instrument, School of Instrument Science and Optoelectronics Engineering, Hefei University of Technology, Hefei 230009 Anhui, China*

[3]*Institute for Translational Brain Research, MOE Frontiers Center for Brain Science, Fudan University, Shanghai 200032, China*

[4]*Department of Physics, Faculty of Science, University of Zagreb, Bijenička c. 32, Zagreb 10000, Croatia*

[5]*Collaborative Innovation Center of Extreme Optics, Shanxi University, Taiyuan, Shanxi 030006, China*

*\*e-mail: tanya@nankai.edu.cn; zgchen@nankai.edu.cn*

## Abstract:

**Optical forces - studied since the earliest days of laser physics - continue to reveal rich dynamics and enable powerful tools for manipulation of objects on micro- and nanoscales, and even individual atoms. Lateral optical forces, which act perpendicular to the direction of beam propagation, are particularly intriguing but have largely been restricted to interface geometries such as air–water boundaries. Here, we realize tunable lateral optical force entirely within a fluid environment by using Janus particles: dielectric microspheres half-coated with gold. We show that the lateral optical force arises from scattering asymmetry induced by the asymmetric structure of the particles; it can be tuned by adjusting the polarization angle of a linearly polarized beam, but also particle parameters including their size and orientation. Experimentally, we directly observe fully reversible lateral propulsion of Janus particles in water merely by rotating the polarization direction, in excellent agreement with theoretical predictions. These results establish a new mechanism for programmable, polarization-controlled optical manipulation, with promising implications for biophotonics, microfluidics, and active soft-matter systems.**

**Introduction**

Optical forces (OFs), arising from the transfer of linear and angular momentum of light to matter, have been widely exploited for non-contact optical manipulation [1-8]. Beyond the extensively studied longitudinal OFs, which act parallel to and depend strongly on the illumination direction [9-12], lateral optical forces (LOFs) - acting perpendicular to the direction of incidence - have emerged as an intriguing research frontier owing to their rich physical origin and broad application potential [13-28].

The fundamental origin of LOFs lies in symmetry breaking of the light scattering, which is induced by the optical field itself or the scattering environment, which leads to a net lateral force [18, 20]. Early demonstrations of LOFs relied on illuminating geometrically asymmetric structures, such as semicylindrical rods, with linearly polarized plane waves [13, 25]. Subsequent approaches have expanded this concept by breaking the scattering symmetry through interfaces using chiral or achiral particles [23, 24, 29, 30], exploiting multiple scattering between particles and substrate [14], leveraging spin-orbit coupling to excite plasmonic resonances in substrate [16], and utilizing multipole interactions to induce lateral asymmetry of the scattered light field [22]. Furthermore, Belinfante's spin momentum - a component of the Poynting vector proportional to the curl of the spin angular momentum [31] - can generate LOFs through spin-momentum inhomogeneities in structured optical fields, including interferometric configurations [32], evanescent waves [15, 31], tightly focused beams [33, 34], and complex structured light fields [35, 36]. More recently, LOFs have also been demonstrated in achiral scatterers of arbitrary size, generated by the interference between electric and magnetic multipoles within the particles with broken fundamental (electric-magnetic) symmetry in the material itself [18].

For and individual homogeneous particle, an interface is required to generate a lateral force to eliminate the uncertainty of LOF magnitude. However, interface can also generate uncertain factors. For example, an LOF is much smaller than the other forms of optical force [17, 30], thus posing a challenge to the optical measurement and utilization of LOFs. Interface-free LOFs have become a prominent research focus, as they offer the potential for far greater flexibility in optical manipulation. Theoretically, complex structured light can achieve lateral forces without interfaces [18, 32]. For instance, LOFs without interfaces can be induced by spin angular momentum inhomogeneities [21], or generated through multipole interactions within a single nanowire [22]. However, their experimental realizations are nontrivial and extremely challenging. Recently, the photon-recoil-based manipulation technique has been employed for realization of optical forces by shaping the imaginary Poynting momentum of a flat-top beam, circumventing the requirement for intensity or phase variations along the manipulation pathway [37]. However, the demonstration of LOFs on nanoparticles under illumination by a single linearly polarized beam still remains a challenge in the absence of an interface.

In practical applications, interface-free control can provide greater flexibility and offer broader applicability. Thus, here we exploit the intrinsic asymmetry of Janus particles suspended in water to generate LOFs without relying on interfaces. Janus particles are broadly defined as composite micron- or nanoscale artificial objects with distinct chemical or physical properties on opposite sides [38, 39]; they are typically referred to as two-faced particles. Due to their unique asymmetric structure and multifunctionality, Janus particles have found widespread use in biological systems and have also attracted significant attention in the field of optical manipulation [40-49]. For instance, Janus particles with a gold cap can be optically rotated [42, 43], directionally steered [46-48], or accelerated [49] by tailored light fields, enabling rich light-controlled dynamics and opening new possibilities for advanced optical manipulation applications.

In this work, we introduce an interface-free strategy for generating switchable LOFs by exploiting geometry-induced symmetry breaking in particles. Specifically, we demonstrate a simple yet effective approach to produce controllable LOFs on Janus particles (polystyrene microspheres partially coated with a gold layer) suspended in water using only linearly polarized light. We show that the structural asymmetry of the particles breaks the lateral symmetry of the scattering patterns in the surrounding fluid, thereby producing measurable LOFs. The resulting forces depend critically on both the incident angle and the polarization angle of the illumination (approximately a plane wave). Systematic numerical studies further reveal that the LOF magnitude is governed by key particle parameters including the gold-layer orientation, particle radius, and coating thickness. Experimentally, we demonstrate the polarization-dependent switching of LOFs through observing real-time control of particle motion in water. Our results provide new insight into polarization-driven optical manipulation and offer a promising toolset for interface-free control of asymmetric particles, with potential applications in biophotonics and the manipulation of irregular biological matter.

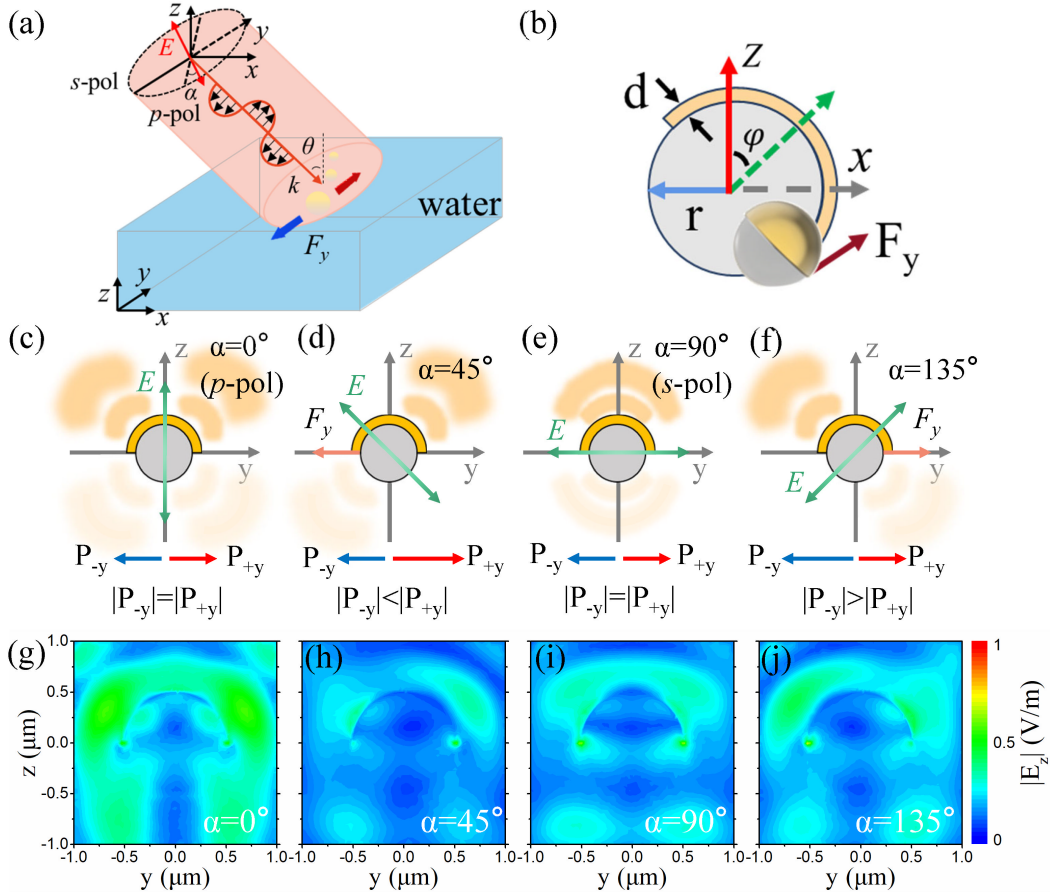

**FIG. 1. Schematic illustration of lateral optical forces (LOFs) on a Janus nanoparticle in an aqueous environment under linearly polarized illumination.** (a) Schematic diagram of the LOFs acting on Janus particles immersed in water. A linearly polarized plane wave with a wavelength of 1064 nm and a polarization angle of $\alpha$ is incident at an angle $\theta$ onto Janus particles immersed in water. (b) A Janus particle consists of a polystyrene bead half-coated with a thin gold layer. The geometrical parameters $\varphi$ (dashed green arrow) and $d$ describe the orientation and thickness of the gold layer, respectively, while $r$ (blue solid arrow) is the radius of the polystyrene sphere. The rose-red arrow indicates the direction of the lateral optical forces $F_y$. (c) and (e) At polarization angles $\alpha = 0°$ ($p$-polarized) and $\alpha = 90°$ ($s$-polarized), scattering

in the $y$-direction remains symmetric, therefore the lateral optical forces $F_y$ vanish. Unless explicitly stated, the orientation of the Janus particle with the gold layer is assumed to have the gold coating oriented at $\varphi = 0°$, i.e., the coating apex aligned with the z axis, and the wavelength of the incident optical field is fixed at 1064 nm. (d) and (f) Under diagonal polarization ($\alpha = 45°$ and $\alpha = 135°$), the scattering symmetry in the $y$-direction relative to the $x$-$z$ plane is broken. When the polarization angle $\alpha = 45°$, the lateral scattered light momentum $P_{-y}$ in the $-y$ direction is smaller than the lateral scattered light momentum $P_{+y}$ in the $+y$ direction, which leads to the lateral optical force $F_y$ along the $-y$ direction. The opposite occurs for the polarization angle $\alpha = 135°$. (g)-(j) The near-field scattering $|E_z|$ in the $y$-$z$ plane ($x = 0$) at incidence polarization angles $\alpha = 0°$, 45°, 90°, and 135°, respectively. The Janus particle is a polystyrene bead ($n_{ps} = 1.6$) with a radius of 500 nm, and is half-coated with a gold layer ($n_{gold} = 0.28 + 6.4i$) of thickness 20 nm. The background environment is water ($n_b = 1.33$), and the incident angle is $\theta = 60°$.

**Results and Discussions**

**1. Asymmetric scattering of a linearly polarized beam from a Janus particle**

The schematic diagram of the LOFs on a Janus nanoparticle in an aqueous environment under linearly polarized illumination is shown in Fig. 1(a). A plane wave with wavelength $\lambda = 1064$ nm is incident on the Janus particle at an angle $\theta$ relative to the z-axis. The polarization angle $\alpha$ is defined as the rotation angle relative to $p$-polarization. Figure 1(b) shows a schematic diagram of the Janus particle model: a polystyrene bead half-coated with a thin gold layer. According to the Drude model [50, 51], the complex refractive index of the gold layer is $n_{gold} = 0.28 + 6.4i$, calculated from the frequency-dependent dielectric permittivity $\varepsilon(\omega)$ via $\varepsilon(\omega) = \varepsilon_\infty - \frac{\omega_p^2}{\omega^2 + i\omega\gamma}$, where $\varepsilon_\infty$ represents the high-frequency limit dielectric permittivity, $\gamma$ is the damping constant, and $\omega_p$ is the plasma resonance frequency. The parameters used in our simulations are $\varepsilon_\infty = 8.7499$, $\gamma = 0.0691$ eV and $\omega_p = 9.0146$ eV [52].

As schematically illustrated in Figs. 1(c) and 1(e), for polarization angles $\alpha = 0°$ or 90°, the momentum of scattered light remains symmetric relative to the $x$-$z$ plane. In this case, no net lateral momentum is transferred to the particle, and the LOFs vanish. However, under the diagonal polarization (polarization angles $\alpha = 45°$ and 135°, in Figs. 1(d) and 1(f)), the symmetry of light scattering relative to the $x$-$z$ plane is broken. When the polarization angle is $\alpha = 45°$, the magnitude of the lateral scattered momentum $P_{-y}$ is smaller than that for the lateral scattered momentum $P_{+y}$ in the $y$ direction. Consequently, the lateral momentum conservation results in a net LOF acting in the $-y$ direction on the Janus particle. When $\alpha = $ 135°, the imbalance reverses. This establishes that LOFs in water can be regulated solely by tuning the polarization angle of the incident wave.

To examine the relationship between polarization and LOFs, we performed COMSOL Multiphysics simulations (see Materials and Methods) of the near-field scattering of Janus particles in the $y$-$z$ plane. A linearly polarized plane wave at $\theta = 60°$ was used, and $\alpha$ was varied at values 0°, 45°, 90°, and 135°. Figures 1(g)-1(j) display the corresponding near-field distributions of $|E_z|$. Pronounced symmetry breaking in the scattered field is observed for diagonal polarization, indicating a breakdown of transverse momentum symmetry and hence a net transfer of lateral momentum to the particle. This mechanism is evident in Figs. 1(h) and 1(j), where $\alpha = 45°$ or 135° yields asymmetric scattering. In contrast, for $\alpha = 0°$ and 90° (Figs. 1(g) and 1(i)), the scattering remains symmetric and no LOF is generated. Because this mechanism relies on particle geometry rather than an interface, it enables a fully bulk (interface-free) route to LOF generation.

To further clarify the physical origin of polarization-dependent LOFs, we computed the far-field scattering distribution under linearly polarized illumination ($\theta = 60°$). Figure 2 shows that the far-field intensity exhibits strong symmetry breaking along the $y$-direction for $\alpha = 45°$ and 135°. For $\alpha = 45°$, the scattered momentum toward $+y$ exceeds that toward $-y$, resulting in an LOF in the $-y$ direction ($F_y < 0$). For $\alpha = 135°$, the imbalance reverses ($F_y > 0$). In contrast, $\alpha = 0°$ and 90° produce symmetric scattering and zero LOF. These results confirm that the intrinsic geometric asymmetry of Janus particles is sufficient to generate and switch LOFs in a homogeneous medium, with the polarization angle providing a simple and effective control knob.

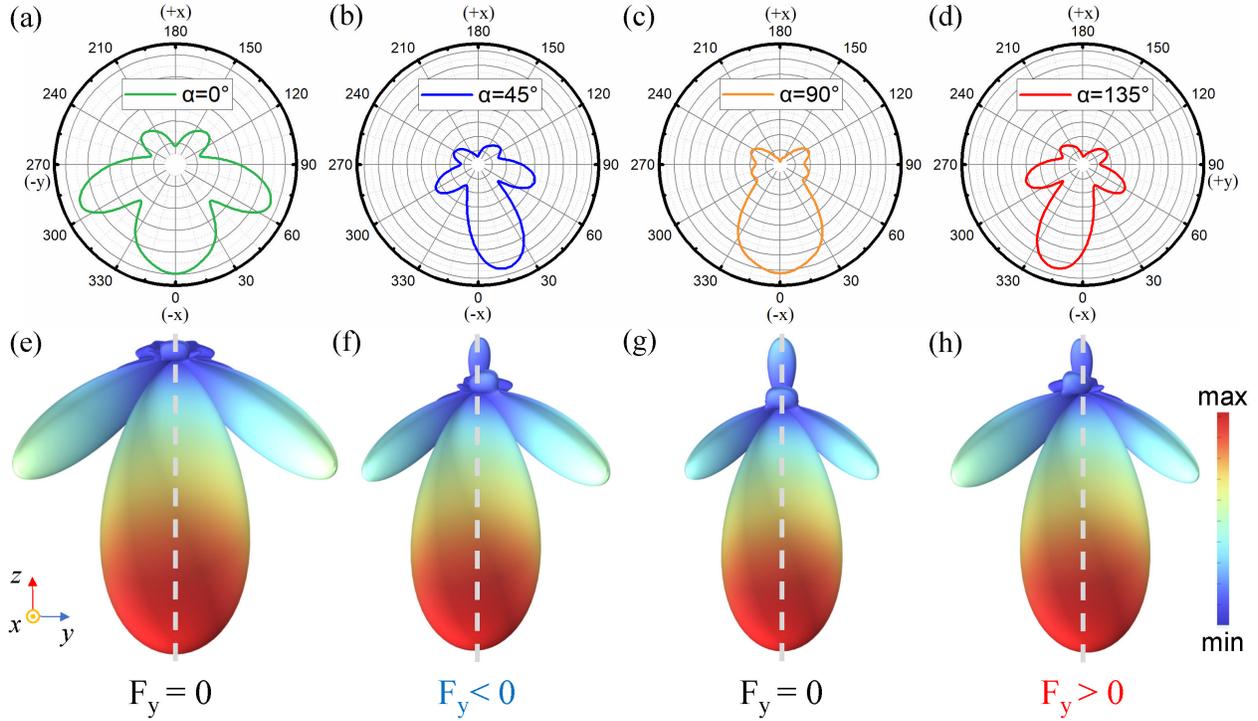

**FIG. 2. Scattering patterns of a Janus particle suspended in water.** The particle is illuminated with a linearly polarized light beam (incident angle $\theta = 60°$) at varying polarization angles $\alpha$. (a-d) Two-dimensional scattering energy profiles in the $x$-$y$ plane for $\alpha = 0°$, 45°, 90°, and 135°. (e–h) Corresponding three-dimensional scattering pattern for better visualization. All other parameters are the same as in Fig. 1.

## 2. Dependence of LOFs on a Janus particle on the parameters

In this section we systematically analyze the dependence of the optical lateral force $F_y$ on the polarization angles $\alpha$ and the incident angle $\theta$ under linearly polarized illumination. The results correspond to a representative Janus particle consisting of a 500-nm-radius polystyrene sphere half-coated with a 20-nm-thick gold layer. Figure 3 shows the LOFs acting on such Janus particle in water. As shown, the LOF induced by a linearly polarized plane wave can be effectively tuned by varying the polarization angle $\alpha$. The force switches direction with $\alpha$, reaching its maximum magnitude ($\sim 1.1\,\text{pN}/(\text{mW}\,\mu\text{m}^{-2})$) at $\alpha = 135°$ and its minimum at $\alpha = 45°$.

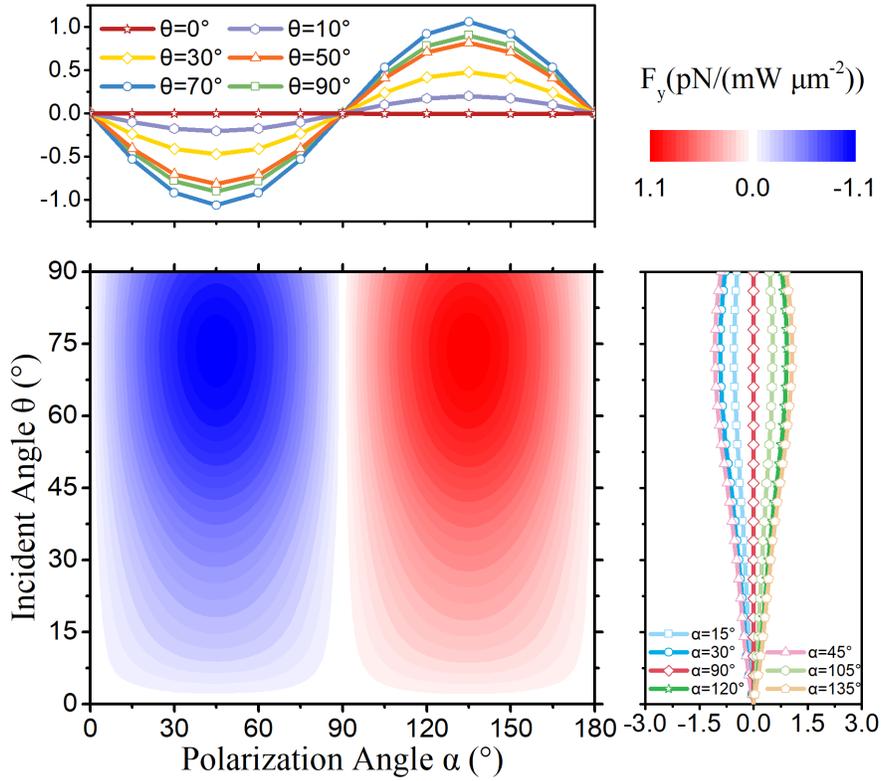

**FIG. 3. Calculated force map showing the lateral force component $F_y$**. The magnitude of the LOF on a Janus nanoparticle in water is plotted as a function of the polarization angle $\alpha$ and incident angle $\theta$. The Janus particle used here is a 500 nm polystyrene sphere half-coated with a gold layer of thickness 20 nm.

To further examine how particle geometry affects LOFs, we numerically studied the influence of three key parameters: gold layer orientation, gold layer thickness, and particle size (see Fig. 4). In Figure 4(a), for a fixed incident angle $\theta = 60°$, the LOFs remain tunable via the polarization angle $\alpha$, but their sign and magnitude depend strongly on the orientation of the gold layer. The orientation defines the particle's geometric asymmetry, which in turn dictates the asymmetry of the scattered light field and the resulting LOF. Notably, when the polarization angle is set to $\alpha = 45°$ and $135°$, the LOFs reverses direction as the gold layer orientation as $\varphi$ crosses roughly $141°$. In contrast, when $\alpha = 0°$ or $90°$, the LOFs remains zero regardless of orientation - consistent with the symmetric - scattering cases illustrated in Fig. 2.

Next, we investigate the dependence of the LOF magnitude on shell thickness and particle size. Figure 4(b) indicates that LOF magnitude varies significantly with both $\varphi$ and $\theta$. Figure 4(c) shows that, with the fixed parameters $r = 500$ nm, $\theta = 60°$ and $\varphi = 0°$, the LOF increases with the thickness $d$ and saturates beyond ~ 20 nm, reflecting the skin-depth-limited contribution of the metallic layer. In contrast, Fig. 4(d) reveals a strong positive correlation between the polystyrene radius $r$ and the LOF when the coating thickness is fixed at 20 nm and orientation at $\varphi = 0°$. This indicates that particle size - rather than coating thickness - plays the dominant role in enhancing lateral scattering asymmetry.

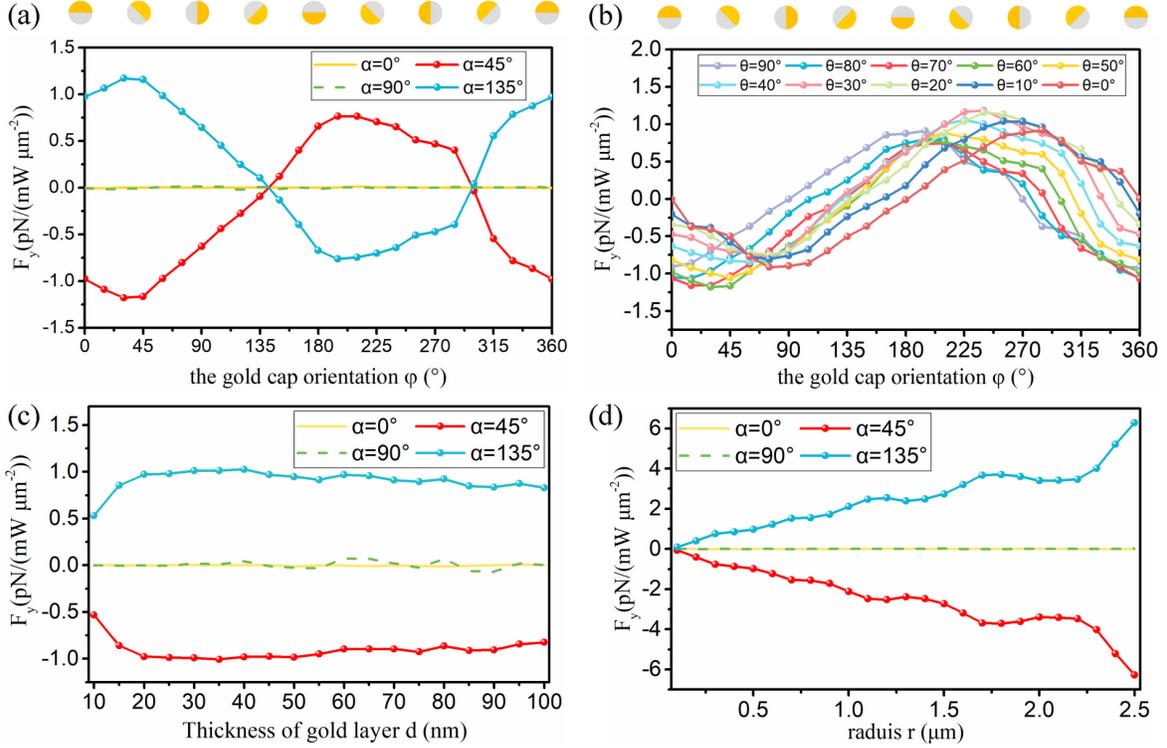

**FIG. 4. Effect of Janus particle parameters on switchable LOFs.** (a-b) LOFs for different orientations of the gold layers: the geometry of the Janus particle is consistent with that in Fig. 3. (a) The rotation angle $\varphi$ of the gold layer varies from 0 to 180 degrees and the incident angle keeps at $\theta = 60°$. (b) The relationship between LOFs and the incidence angle $\theta$ of a plane wave with polarization angle $\alpha = 45°$ and the rotation angle $\varphi$ of the gold layer. (c, d) The influence of Janus particle size on LOFs when the rotation angle $\varphi$ is fixed at 0° and the incident angle at $\theta = 60°$. The variation of the LOFs (c) with the polystyrene radius $r = 500$ nm, the gold layer thickness $d$, and (d) with the polystyrene radius $r$ when the gold layer thickness is $d = 20$ nm.

### 3. Experimental observation of switchable LOFs

To directly observe the switchable lateral force, a linearly polarized beam illuminates Janus particles suspended in water at an incident angle of $\theta = 60°$. A schematic of the setup is shown in Figure 5(a). The laser beam ($\lambda$ = 1064 nm, maximum power of 1.5 W) first passes through an optical beam expansion system composed of two lenses with focal lengths of 15 mm and 200 mm, respectively, to enlarge the beam and improve its spatial focusing quality. It is then collimated by a 400-mm lens to produce an illumination spot of approximately 400 μm. A half-wave plate is used to rotate the polarization angle. The sample is illuminated from below with a non-coherent white-light LED, and particle motion is imaged through a 10× microscope objective (NA = 0.3). A 1064-nm notch filter blocks scattered laser light, and trajectories are recorded with a CCD camera.

In our experiment, the Janus particles consist of polystyrene spheres with a diameter of 5 μm half-coated with a 15 nm gold layer (see Materials and Methods). Due to the influence of particle gravity, the gold cap naturally orients downward, corresponding to a rotation angle of $\varphi$ = 180°. Figure 5(b) shows the simulated LOFs for this geometry under illumination by a linearly polarized plane wave. To account for fabrication-induced variations, we computed the forces for gold thicknesses of 10, 15, and 20 nm. The direction of the LOFs remained unchanged across this range. Figure 5(c) shows the measured particle trajectories recorded over 4 minutes. Because the incident beam exerts a constant radiation-pressure force in the +x direction, any LOF causes the trajectory to tilt in the y-direction. By rotating the half-wave plate, we continuously tune the sign and magnitude of the LOF. For $\alpha$ = 45°, the particles drift toward the -y direction (Supplementary Movie 1) for about 10 μm, whereas for $\alpha$ = 135°, they drift toward the +y direction (Supplementary Movie 2) for about 13 μm. These experimental observations are in excellent agreement with our simulation, confirming that polarization alone can reverse the LOF direction.

This interface-free scheme provides a robust and flexible method to manipulate LOFs in aqueous environments, enabling polarization-controlled transport of asymmetric particles.

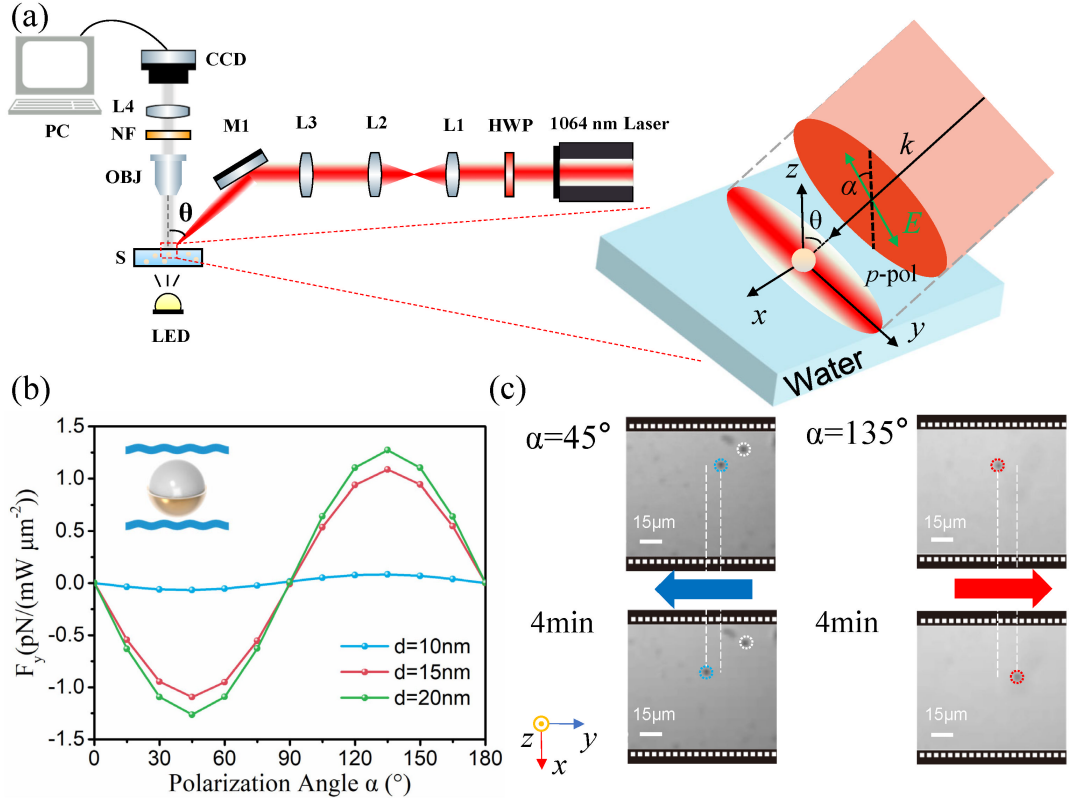

**FIG. 5. Experimental observation of switchable LOFs on Janus particles in water.** (a) Schematic of experimental setup. M: mirror; HWP: half-wave plate (1064 nm); L1: lens 1 ($f' = 15$ mm); L2: lens 2 ($f' = 200$ mm); L3: lens 3 ($f' = 400$ mm); L4: lens 4 ($f' = 200$ mm); S: sample stage; OBJ: ×10 objective; NF: notch filter (1064 nm). (b) Calculated lateral optical forces $F_y$ on Janus particles (radius $r = 2.5$ μm) in water for different gold-coating thicknesses ($d = 10$ nm, 15 nm, 20nm). The inset in the upper-left corner illustrates the gold-layer orientation. (c) Switching of LOF direction on Janus particles by varying the linear polarization angles $\alpha$. Blue and red circles mark particle positions under different polarization angles; the white circle indicates a fixed reference point. Arrows denote the corresponding direction of the LOF. Particle trajectories were recorded over 4 minutes. Scale bar: 15μm.

**Conclusion**

In summary, we have shown that geometrically asymmetric Janus particles can generate measurable LOFs directly within a homogeneous aqueous medium, without relying on interfaces or field gradients. Such LOFs arise from polarization-dependent symmetry breaking in the particles' scattered fields, which produces an imbalance in transverse momentum and thus a net lateral force. By systematically tuning the polarization and incidence angle of the illuminating beam, together with particle-specific parameters such as size, coating thickness, and gold-layer orientation, we demonstrate precise and reversible control over both the sign and magnitude of the LOFs. Our experiments corroborate the theoretical predictions, revealing real-time, polarization-switched lateral propulsion of Janus particles in water. This interface-free mechanism represents a clean and versatile alternative to previously reported LOF-generation strategies that rely on substrate interactions, chiral responses, tightly focused beams, or structured optical fields.

Beyond establishing a new physical scheme for symmetry-mediated LOFs, this work introduces a generalizable optical control scheme for asymmetric microparticles. The demonstrated ability to steer particle motion using only the polarization state of a broad beam - without spatial shaping or complex trapping geometries - offers significant practical advantages. It paves the way for applications in microrobotics, optical sorting, targeted delivery, biophotonics, and reconfigurable lab-on-a-chip platforms, particularly where gentle, interface-free manipulation is required. More broadly, our results suggest that polarization-governed momentum exchange in asymmetric scatterers may provide a powerful and unexplored degree of freedom for next-generation optical micromanipulation technologies.

**Materials and Methods**

**Simulation of light fields and lateral optical forces**

We consider a linearly polarized plane wave, with the propagation vector $\boldsymbol{k}$ lying in the $x$-$z$ plane. The corresponding electric and magnetic field vectors are expressed as

$$\boldsymbol{E} = \left(-\cos\theta \cos\alpha\, \boldsymbol{e_x} + \sin\alpha\, \boldsymbol{e_y} + \sin\theta \cos\alpha\, \boldsymbol{e_z}\right) \exp[i(k_x x + k_z z)]$$

$$\boldsymbol{B} = \frac{1}{c}\left(-\cos\theta \sin\alpha\, \boldsymbol{e_x} - \cos\alpha\, \boldsymbol{e_y} + \sin\theta \sin\alpha\, \boldsymbol{e_z}\right) \exp[i(k_x x + k_z z)], \qquad (1)$$

where we assume the time dependence of $\exp[i\omega t]$. $\boldsymbol{e_x}$, $\boldsymbol{e_y}$ and $\boldsymbol{e_z}$ are the unit vectors in the Cartesian coordinates, $c$ the speed of light in vacuum, $i$ denotes the imaginary unit, and $k_x$ and $k_z$ represent the components of the wave vector $\boldsymbol{k}$ along the $x$ and $z$ directions, respectively. The incidence angle and the polarization angle are defined as $\theta$ (with respect to the $z$-direction) and $\alpha$ (with respect to the $p$-polarization), respectively (Fig. 1).

We employed the finite element method in COMSOL Multiphysics to perform the rigorous simulations of the light fields and lateral forces. By integrating the time-averaged Maxwell stress tensor $\langle \overleftrightarrow{\mathbf{T}} \rangle$ over the entire surface enclosing the Janus particle, the time-averaged optical force $\langle \mathbf{F} \rangle$ is given by [53, 54]

$$\langle \mathbf{F} \rangle = \oint \hat{\mathbf{n}} \cdot \langle \overleftrightarrow{\mathbf{T}} \rangle dS, \tag{2}$$

where

$$\langle \overleftrightarrow{\mathbf{T}} \rangle = \frac{1}{2} \mathrm{Re} \left[ \varepsilon_r \varepsilon_0 \mathbf{E} \mathbf{E}^* + \mu_0 \mathbf{H} \mathbf{H}^* - \frac{1}{2} (\varepsilon_r \varepsilon_0 \mathbf{E} \cdot \mathbf{E}^* + \mu_0 \mathbf{H} \cdot \mathbf{H}^*) \overleftrightarrow{\mathbf{I}} \right], \tag{3}$$

and $\hat{\mathbf{n}}$ is the outward normal unit vector to the surface of integration, $\mathbf{E}$ and $\mathbf{H}$ are the electric and magnetic fields, respectively. $\overleftrightarrow{\mathbf{I}}$ is the 3 × 3 unitary tensor, and $\varepsilon_0$ and $\mu_0$ are the vacuum permittivity and vacuum permeability, respectively. In an aqueous media, the relative dielectric constant is $\varepsilon_r = 1.33^2$. For convenience, the maximum intensity of linearly polarized plane wave is normalized to 1 mWμm$^{-2}$. Due to the nanoscale thickness of the gold layer, the simulation accuracy exhibits highly sensitivity to the mesh size. Finer meshes results in longer calculation time. In order to achieve better force accuracy within a promisingly shorter calculation period, we implemented a careful control over mesh scheme: a maximum element of 5 nm was set at the integration boundary, while the minimum and maximum mesh element sizes for the Janus particles are 10 nm and 20 nm, respectively, and the maximum mesh in the medium is chosen to be $\lambda/(10n)$, where $n$ is the refractive index of the surrounding medium.

**Sample preparation**

The Janus particles used in our experiments were fabricated by coating one hemisphere of 5-μm-diameter polystyrene microspheres with a thin gold layer via magnetron sputtering. Briefly, a droplet of the polystyrene microsphere suspension was first deposited onto a clean, hydrophilic glass substrate, where the particles self-assembled into a monolayer upon drying. A gold film with a nominal thickness of 15 nm was then deposited onto the exposed hemisphere of the microspheres at a rate of 4 nm/min using magnetron sputtering. To retrieve the Janus particles, the coated substrate was immersed in deionized water and sonicated in an ultrasonic bath, allowing the particles to detach and disperse into a stable aqueous suspension. The resulting suspension was subsequently centrifuged to remove impurities and then stored at low temperature until use.

## Acknowledgments

This research was supported by the National Key R&D Program of China under Grant No. 2022YFA1404800, the National Natural Science Foundation (W2541003, 12134006, 12374309, 11922408, 62475047), the 111 Project of B23045, and the project "Implementation of cutting-edge research and its application as part of the Scientific Center of Excellence for Quantum and Complex Systems, and Representations of Lie Algebras", Grant No. PK.1.1.10.0004, co-financed by the European Union through the European Regional Development Fund—Competitiveness and Cohesion Programme 2021-2027.

## Conflict of Interest

The authors declare no conflict of interest.

## Date Availability

All source data that support the plots within this paper and other findings of this study are available from the corresponding authors upon reasonable request.